\documentclass{emulateapj}

\shorttitle{Satellites in the Millennium simulation}
\shortauthors{V. Quilis \& I. Trujillo} 

\def\gsim{ \lower .75ex \hbox{$\sim$} \llap{\raise .27ex \hbox{$>$}} }
\def\lsim{ \lower .75ex \hbox{$\sim$} \llap{\raise .27ex \hbox{$<$}} }

\begin{document}
\title{Satellites around massive galaxies since z$\sim$2: confronting the Millennium simulation  with observations}

\author{Vicent Quilis} 
\affil{Departament d'Astronomia i Astrof\'{\i}sica, 
Universitat de Val\`encia, 46100 - Burjassot, Val\`encia, Spain}
\email{vicent.quilis@uv.es}
\and
\author{Ignacio Trujillo\altaffilmark{1,2}} 
\affil{Instituto de Astrof\'{\i}sica de Canarias, 
c/ V\'{\i}a L\'actea s/n, E38205 - La Laguna, Tenerife, Spain}
\altaffiltext{1}{Ram\'on y Cajal Fellow}
\altaffiltext{2}{Departamento de Astrof\'isica,
Universidad de La Laguna, E-38205 La Laguna, Tenerife, Spain}

\begin{abstract}

 Minor merging has been postulated as the most likely evolutionary path to produce the increase in size and mass observed in the massive galaxies since z$\sim$2. In this Letter, we test directly this hypothesis comparing the population of satellites around massive galaxies in cosmological simulations versus the observations.  We use state-of-the-art, publically
available, Millennium I and II simulations and the associated semi-analytical galaxy catalogues
to explore the time evolution of the fraction of massive galaxies that 
have satellites, the number of satellites per galaxy,
the projected distance at which the  satellite locate from the host galaxy, and the
mass ratio between the host galaxies and their satellites. The three virtual galaxy catalogues considered here,  
overproduce the fraction of galaxies with 
satellites by a factor ranging between 1.5  and 6 depending on the epoch, whereas the mean 
projected distance and ratio of the satellite mass over host mass 
are in closer agreement with data. The larger pull of satellites in the semi-analytical samples
 could suggest that the size evolution found in previous hydrodynamical simulations  is an artifact due to the larger number of infalling satellites compared to the real Universe. 
These results advise to revise the physical ingredients implemented in the semi-analytical models in order to reconcile 
the observed and 
computed fraction of galaxies 
with satellites, and eventually, it
would leave some room to other mechanisms  explaining the galaxy size growth not related to the minor merging.

\end{abstract}


\section{Introduction}

Accretion of minor satellites has been postulated as the most likely mechanism to explain the significant size
evolution (Daddi et al. 2005; Trujillo et al. 2006) of the massive galaxies over cosmic time. This idea fits with
many observational indirect evidence: the progressive growth of the wings of the profiles of the massive galaxies
with time (Bezanson et al. 2009; Hopkins et al. 2009, van Dokkum et al. 2010, Carrasco et al. 2010), the ($\sim$1.5) larger velocity dispersion of the massive galaxies at high-z compared to present-day equally massive objects  (e.g. Cenarro \& Trujillo 2009, Cappellari et al. 2009,
Onodera et al. 2010, Newman et al. 2010, van de Sande et al. 2011), and  the
expected mass growth by a factor of two of the massive galaxies with time suggested by the observations (see
e.g. Trujillo et al. 2011).

On the theoretical side, Naab et  al. (2009)  conducted a pioneer  work on exploring  the minor merging effect 
on a cosmologically  motivated evolution of  a massive galaxy  since z$\sim$3.  Their  study supported  the  idea
that  minor merging  can explain  simultaneously the  size and  velocity dispersion evolution, as well as a moderate increase of the stellar mass by a factor of two. This work has been
later confirmed by many other authors (e.g. Sommer-Larsen \& Toft 2010, Feldmann et al. 2010, Oser et al. 2012).
Nonetheless, although both the observational and theoretical side seem to converge in a unique solution to the
problem of the size evolution of the massive galaxies, a direct confrontation of both the theory and the
observations has not yet been conducted. Inferring the merger rate (e.g. L\'opez-Sanjuan et al. 2011) from the observations is not straightforward due to the large uncertainties in the determination of the merging time scales. Alternatively,
 an immediate way of testing the model with the data is counting
 the number of satellites around the massive galaxies. These objects will likely be involved in the growth of the massive
objects. Observations have matured enough to allow a robust determination of the fraction of massive galaxies with
satellites at different redshifts and for different mass ratios (e.g. Newman et al. 2012, M\'armol-Queralt\'o et al. 2012,
L\'opez-San Juan et al. 2012). On the other hand, present cosmological simulations are large enough to permit the estimation of this fraction with accuracy.

In this Letter, we explore the model predictions about the properties of the satellites surrounding massive
galaxies and their evolution with cosmic time. In particular, we compare the changes with redshift of the fraction
of satellites, radial projected distances and stellar mass ratio of three different semi-analytical models (Bower
et al. 2006, DeLucia et al. 2007, Guo et al. 2011) based on the Millennium simulation with the data taken by 
M\'armol-Queralt\'o et al. (2012).We find that although the general trends with redshift are reproduced (i.e. the
fraction of massive galaxies with satellites only changes moderately with cosmic time in the simulations), all the 
models tend to overpredict this fraction.

\section{Galaxy catalogues}

We use the public release of  two very large N-body
simulations Millennium I~\citep{springel05} and
Millennium-II~\citep{bk09}. The cosmology assumed in
both simulations is a $\Lambda CDM$ with the following
parameters: $\Omega_m=0.25$, $\Omega_b=0.045$, $\Omega_{\Lambda}=0.75$,
$n=1$, $\sigma_8=0.9$, $H_o=73 \,km s^{-1}Mpc^{-1}$. The two simulations use
the same number of particles, $2160^3$, but cover different volumes and, therefore,
they have different numerical resolutions. Thus, the computational boxes have sides of
$685\,Mpc$ and $137 \,Mpc$, and particles masses of $1.18\times10^9\,M_{\odot}$ and
 $9.45\times10^6\,M_{\odot}$ for the Millennium I and II simulations, respectively.

The dark matter  merger trees are extracted from the simulations using a combination
of friends-of-friends (FoF)~\citep{davis85} and SUBFIND~\citep{springel01} halo finders.
The dark matter halos are converted into galaxies according to different
semi-analytical models which can differ in the phenomenological recipes to introduce
the gas and stellar components in such haloes. In this letter, we consider three
semi-analytical models available in the Millennium database web ~\citep{lemson06}.

The first one is a version of the Durham semi-analytical model \citep{bower06} that
implements feedback due to active galactic nuclei (AGN) as a manner to stop or delay the
formation of cooling flows.
The other two models are the ones by \cite{delucia07} and \cite{guo11}. Both are very similar
being the second one an improvement of the earlier version that modifies or extends some of
the semi-analytical recipes. The new features implemented in the model by \cite{guo11} are:
the separate evolution of sizes and orientations of gaseous and stellar discs, the
size evolution of spheroids, tidal and ram-pressure stripping of satellite galaxies, and the disruption
of galaxies to produce intracluster light.  The effect of AGN feedback was already included in the
earlier version by \cite{delucia07}. The initial mass function (IMF) used to estimate the stellar masses of the simulated galaxies were: Chabrier (2003) for  \cite{guo11} and \cite{delucia07} and Kennicutt (1983) by  \cite{bower06}.

The models by \cite{bower06} and \cite{delucia07} use the Millennium I simulation, whereas
the third model by ~\cite{guo11} is based on the higher resolution
Millennium-II simulation.

\begin{figure*}
\includegraphics[scale=.8]{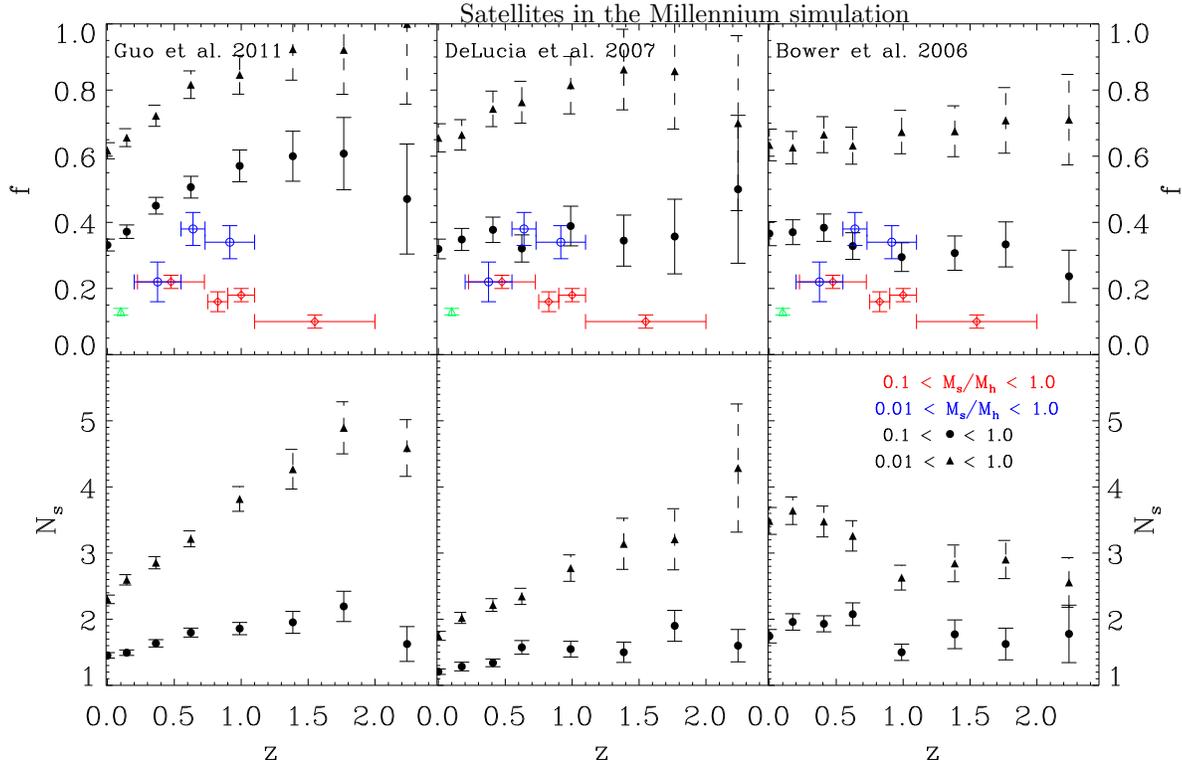}
\caption{The columns stand for the results of three galaxy catalogues based on 
different semi-analytical models. For each model, and from top to bottom: the 
fraction of massive galaxies that have at least one satellite within a sphere
of $100\,kpc$ radius and a projected distance smaller than $100\,kpc$, and the average number of satellites 
per massive galaxy when they have one of such objects around.
The full circles (triangle) stand for the 
satellites with stellar mass ratios of  $0.1< M_{s}/M_{h}< 1$ ($0.01< M_{s}/M_{h}< 1$).
The error bars represent one standard deviation. The observational data 
from \cite{esther12} are overploted as red (blue) 
open circles (diamonds) for mass ratios of $0.1< M_{s}/M_{h}< 1.0$ 
($0.01< M_{s}/M_{h}< 1$). The local observational reference (z=0.1) from Liu et al. (2011)
for the fraction of  massive galaxies with satellites with mass ratios of $0.1< M_{s}/M_{h}< 1$ 
is plotted as a green open triangle, no data are available for smaller satellites.}
\end{figure*}

\begin{figure*}
\includegraphics[scale=.8]{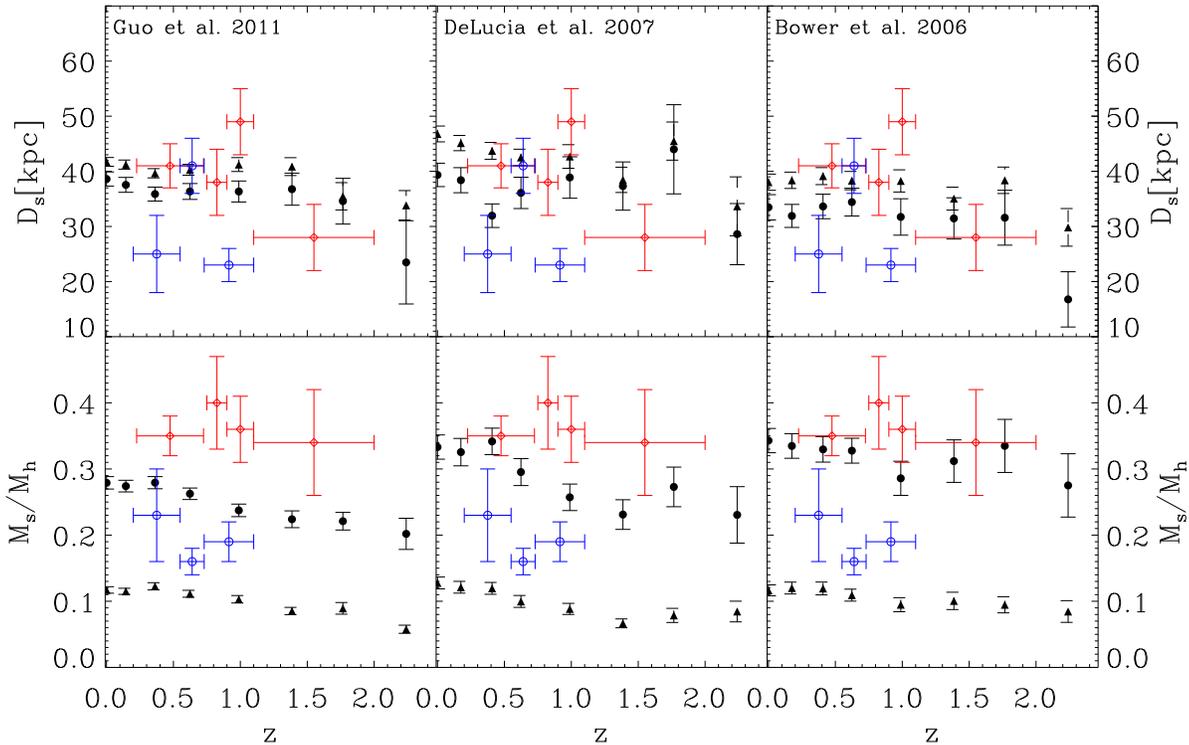}
\caption{From top to bottom, the mean projected
distance of the satellites to their host galaxy and the mean ratio of the satellite stellar mass
over the host stellar mass for the three semi-analytical samples. The symbols and colors 
are defined as in the Fig.1.} 
\end{figure*}

\section{Results}

We have generated three galaxies catalogues using the 
Millennium web application. Each catalogue corresponds to one different semi-analytical
model as previously discussed. For each of the catalogues, we select all the 
massive galaxies -- that we call host galaxies -- as those objects with stellar masses, $M_h$, 
between $10^{11}\,M_{\odot}$ 
and $10^{13}\,M_{\odot}$. 
For each host galaxy, and for the sake of comparison with the data by \cite{esther12},  we
consider a spherical region of radius of $100\,kpc$ (physical units). Within this region, we compute   
the number of satellites with stellar masses  $0.1< M_{s}/M_{h}< 1$ and projected distances smaller than
$100\,kpc$. Once the satellites fulfilling such 
mass ratio and separation conditions are identified,  we calculate the projected distance of
each satellite and  the mass ratio respect to their host galaxy. 
In order to study different mass scales, this process is repeated  for smaller satellites 
with stellar masses in the range of  $0.01< M_{s}/M_{h}< 1$.
Possible bias effects due to the use of different host masses or galaxy type-dependent number of 
satellites are minimized by using the same mass criteria for simulated and observed galaxies 
and by comparing both samples globally, without distinguishing between galaxy types.

In the Figs. 1 and 2, we present the results of the previous analysis on the three considered 
galaxy catalogues. The figures are organized in three columns, each one showing 
the results of one particular catalogue. For all the columns in the Fig.1 , and from top to bottom, 
we display the fraction ($f$) of massive galaxies that have at least one satellite and 
the mean number of satellites ($N_s$) for those host galaxies that have them.
The average 
projected distance of the satellites ($D_s$) and the average ratio of the satellite stellar mass 
over the host galaxy stellar mass ($M_{s}/M_{h}$) are presented in the Fig.2 organized as in Fig.1.
The full circles (triangle) stand for the 
simulated satellites with mass ratios of  $0.1< M_{s}/M_{h}< 1$ ($0.01< M_{s}/M_{h}< 1$).
The error bars are computed as one standard deviation. The observational 
data from \cite{esther12} are overploted as red (blue) 
open circles (diamonds) for mass ratios of $0.1< M_{s}/M_{h}< 1$ 
($0.01< M_{s}/M_{h}< 1$) in both figures. 
The IMF used in this work was Chabrier (2003). These data are representative of the results found 
for different authors (see e.g. Fig. 6 of L\'opez-Sanjuan et al. 2012), and for this reason 
we use it here as a basis of comparison with the simulations.

The Fig. 1 clearly shows that all the semi-analytical models explored in this paper systematically 
overpredict at every redshift the fraction of massive galaxies having satellites in their vicinity. The model expectation is  
larger than what it is currently measured by a factor ranging between 1.5 and 6 depending on the epoch. This factor is independent on the mass ratio considered to probe the satellite population (either 1:10 or 1:100). 

The average projected distance of the satellites to the host galaxy 
in Fig.2 is well reproduced (within the errors) for the case of 1:10 ratio: $\sim 40 \,kpc $. However, in the case of 1:100 ratio, the observed satellites are closer to the massive 
objects than what simulations predicts. Finally, in relation to the typical average mass of the satellites within each mass ratio 
displayed in Fig.2, the models are closer to the observations, particularly in the case of Bower et al. (2006). Another important aspect 
of the observations which agree broadly with the models is the fact that all the properties of the satellites explored here: the fraction of massive galaxies with satellites, their average distance to the host object and the average mass ratio of the satellite depend weakly on the redshift.

Summarizing, model predictions agree qualitatively well with the observations except for the fact that they overpredict on average 
the fraction of massive galaxies with satellites around them by a factor of two.

\section{Discussion}

The most striking of our results is the overabundace of simulated satellites around massive
galaxies by a factor between 1.5 and 6 compared to the observations. At z=0 this is quite surprising as the stellar mass function of the Millennium galaxies 
 (Guo et al. 2011) is perfectly compatible with the observations (P\'erez-Gonz\'alez et al. 2008). Guo et al. (2011) uses the same observational database than in M\'armol-Queralt\'o et al. (2012), so the origin of this discrepancy can not be explained as a consequence of a different dataset. It is also irrelevant the slightly high power spectrum normalization in Millennium 
 ($\sigma_8=0.9$), as no important effects are expected at the scales studied in this work (see  for instance Zentner \& Bullock 2003). 
 The only way to reconcile both results is considering that the spatial 
 distribution of less massive galaxies is not the same in the real and in the virtual Universe. In particular, the satellite galaxies are more clustered (a factor of two on average) around the massive galaxies in the 
 semi-analytical samples than what the observations suggest. It is 
 worthy to note that at $z>0$ the semi-analytical models considered here do not match the observed mass functions, and indeed, they 
 overproduce the number of low mass galaxies.

The fact that semi-analytical satellites appears to be on average a factor of two more common around the massive galaxies than in the observations could point out to a factor of two larger merger time scale in the models than in the real Universe (a scenario already discussed in the literature and which have eluded so far a conclusive answer, see e.g. the discussion in L\'opez-Sanjuan et al. 2012).  Other possible sources of the discrepancy between the abundance of real and virtual satellite galaxies could be 
related with numerical resolution effects or the use of an unrealistic  modeling of the galaxy
formation processes -- for moderate and small masses -- in the semi-analytical models considered in this
Letter. 
In order to discard possible resolution effects, we have performed an extra analysis using two galaxy catalogues produced by the 
same semi-analytical model described in Guo el al. (2011) on both Millennium I and II simulations. The results are extremely 
similar for all the analysed quantities in both cases. This fact discards uncontrolled effects due to the numerical resolution and stresses the crucial
role of the  considered semi-analytical model. In particular, the different treatment for identifying the  central galaxies and for the gas stripping 
between the models by Guo et al. (2011) and DeLucia et al. (2007) should enhanced the satellite galaxy population in Guo's model 
as it is proved by the results shown in Figure 1.
We note, however, that even with such differences  the three models considered here show a clear excess in the fraction of satellites around massive galaxies.

Another remarkable result of the observations is the constancy of the fraction of massive galaxies with satellites at all redshift. This suggests that the number of satellites per host is in equilibrium leading to a constant accretion of stellar mass by the host galaxy (Nierenberg et al. 2012). This is in contrast with the decline of this fraction since $z\sim 1.2$ in the  higher
resolution Millennium II simulation. We remark, however, that in the lower resolution Millennium I simulations this
 fraction seems also not to change with cosmic time in better agreement with the observations. We can link this finding with the radial distribution of the satellites. In all the three semi-analytical catalogues, the projected average distance of the satellites is almost constant. 
This trend is perfectly consistent with the observations, although the 
actual values for the average projected distance are only similar between data and models for the larger satellites. A constant 
average radial distance of the satellites indicates no significant evolution in the radial profile of satellites, in agreement 
with the result found by Budzynski et al. (2012).

Finally, it is important to discuss the results found here in relation to the role of minor merging in the evolution in mass and size of the massive galaxies as commented in the Introduction. Both, observations and hydrodynamical simulations agree on the important role played by the satellites at feeding the host massive galaxies and producing an unavoidable increase in their sizes and masses. In fact,  such hydrodynamical simulations, as the ones presented in Oser et al. (2012), are able to explain fully the evolution in size of the massive galaxies by the effect of minor merging. However, the results presented in the present work indicate that these findings should be taken with caution, since these simulations do not include AGN feedback and therefore, even a larger number of satellites should be expected. In particular, simulations showing that the full size evolution can be explained by minor merging alone must be  also able to reproduce the number of satellites found observationally. Thus, the larger pull of satellites in the simulations could suggest that size evolution found in the hydrodynamical simulations (although in agreement with the observations) is an artifact due to the larger number of infallen satellites compared to the real Universe. This would leave some room to other mechanisms for galaxy size growth not related to the minor merging. 
However, the ambiguities  in the physical processes modeled in the simulations leads to uncertainties in crucial issues like  the merger time scales or the 
efficiency in the size growth that prevent us to conclude this result with full certainty. 

\acknowledgments This work was  supported by the Spanish Ministerio de
Econom\'{\i}a y Competitividad (MINECO)(grants   AYA2010-21322-C03-01, AYA2010-21322-C03-02  and
CONSOLIDER2007-00050)   and    the   Generalitat   Valenciana   (grant
PROMETEO-2009-103).  The Millennium Simulation databases used in this paper and the web 
application providing online access to them were constructed as part of the activities of the German Astrophysical Virtual Observatory.
The authors thank
the anonymous referee for constructive criticism and E. M\'armol-Queralt\'o for interesting discussions.


\begin{thebibliography}{}

\bibitem[Bezanson et al. (2005)]{bez05} Bezanson, R., van Dokkum, P. G., Tal, T., Marchesini, D., Kriek,
M., Franx, M. \& Coppi, P., 2009,ApJ, 697, 1290

\bibitem[Bower et al. (2006)]{bower06}Bower R. G., Benson A. J., Malbon R., Helly J. C., Frenk C. S., Baugh C.
M., Cole S., Lacey C. G., 2006, MNRAS, 370, 645

\bibitem[Boylan-Kolchin et al. (2009)]{bk09}Boylan-Kolchin M., Springel V., White S. D. M., Jenkins A., Lemson G., 2009, MNRAS, 398, 1150
\bibitem[Budzynski et al. (2012)]{bud12} Budzynski, J.M., Koposov, S., McCarthy, I.G., McGee, S.L., \& 
Belokurov, V., 2012, arXiv:1201.5491

\bibitem[Cappellari et al. (2009)]{cap09} Cappellari, M., et al., 2009, ApJ, 704, L34

\bibitem[Carrasco et al. (2010)]{car10} Carrasco, E. R., Conselice, C. J., Trujillo, I., 2010, MNRAS, 405, 2253

\bibitem[Cenarro et al. (2009)]{cen09} Cenarro, A. J., Trujillo, I., 2009, ApJ, 696, 43

\bibitem[Chabrier (2003)]{cha03} Chabrier G., 2003, PASP, 115, 763

\bibitem[Daddi et al. (2005)]{daddi05} Daddi, E. et al., 2005, ApJ, 626, 680

\bibitem[Davis et al. (1985)]{davis85}Davis M., Efstathiou G., Frenk C. S., White S. D. M., 1985, ApJ, 292,
371

\bibitem[De Lucia \& Blaizot (2007)]{delucia07}De Lucia G., Blaizot J., 2007, MNRAS, 375, 2 
	
\bibitem[Feldmann et al. (2010)]{Feld10} Feldmann, R., Carollo, C. M., Mayer, L., Renzini, A., Lake, G., Quinn, T., Stinson, G. S., Yepes, G., 2010, ApJ, 709, 218

\bibitem[Guo et al. (2011)]{guo11}Guo Q., White S., Boylan-Kolchin M., De Lucia G., Kauffmann G., Lemson G.,  Li C., Springel V., Weinmann S., 2011,MNRAS, 413, 101

\bibitem[Hopkins et al. (2009)]{hopkins09} Hopkins, P. F., Bundy, K., Murray, N., Quataert, E., Lauer, T.
R., Ma, C.-P., 2009, 398, 898

\bibitem[Kennicutt (1983)]{ken83} Kennicutt R. C., 1983, ApJ, 272, 54

\bibitem[Lemson et al. (2006)]{lemson06} Lemson G. and the Virgo Consortium, 2006, astro-ph/0608019

\bibitem[Liu et al. (2011)]{liu11} Liu, L., et al., 2011, ApJ, 733, 62

\bibitem[L\'opez-Sanjuan et al. (2011)]{lop11} L\'opez-Sanjuan, C., et al., 2011, AJ, 530, 20

\bibitem[L\'opez-Sanjuan et al. (2012)]{lop12} L\'opez-Sanjuan, C., et al., 2012, A\&A, submitted, arXiv:1202.4674

\bibitem[M\'armol-Queralt\'o et al. (2012)]{esther12}M\'armol-Queralt\'o E., Trujillo I., P\'erez-Gonz\'alez P. G., Varela J., Barro G., 2012, MNRAS, in press, arXiv:1201:2414

\bibitem[Naab et al. (2009)]{naab09} Naab, T., Johansson, P. H.. Ostriker, J. P., 2009, ApJ, 699, L178

\bibitem[Newman et al. (2010)]{newman10} Newman, A. B.; Ellis, R. S., Treu, T., Bundy, K., 2010, ApJ, 717,
L103

\bibitem[Newman et al. (2012)]{newman12} Newman, A. B., Ellis, R. S., Bundy, K., Treu, T., 2012, ApJ, 746, 612

\bibitem[Nierenberg et al. (2012)]{nier12} Nierenberg, A.M., Auger, M.W., Treu, T., Marshall, P.J., Fassnacht, C.D., 
Busha, M.T.,  2012, arXiv:1202.2125

\bibitem[Onodera et al. (2010)]{Ono10} Onodera, M., et al. 2010, ApJ, 715, L60

\bibitem[Oser et al. (2012)]{Ose12} Oser, L., Naab, T., Ostriker, J. P., Johansson, P. H., 2012, ApJ,
744, 630

\bibitem[Perez-Gonzalez et al. (2008)]{per08} P\'erez-Gonz\'alez, P.G. et al., 2008, ApJ, 675, 234

\bibitem[Sommer-Larsen et al. (2010)]{som10} Sommer-Larsen, J. \& Toft, S., 2010, ApJ, 721, 1755

\bibitem[Springel et al. (2001)]{springel01} Springel V., White S. D. M., Tormen G., Kauffmann G., 2001, MNRAS, 328, 726

\bibitem[Springel et al. (2005)]{springel05}Springel V. et al., 2005, Nat, 435, 629

\bibitem[Springel et al. (2008)]{springel08}Springel V. et al., 2008, MNRAS, 391, 1685

\bibitem[Trujillo et al. (2006)]{trj06} Trujillo I., et al., 2006, MNRAS, 373, L36

\bibitem[Trujillo et al. (2011)]{trj11} Trujillo, I., Ferreras, I., de La Rosa, I. G., 2011, MNRAS, 415,
3903

\bibitem[van de Sande et al. (2011)]{vsa11} van de Sande, J., et al., 2011, ApJ, 736, L9

\bibitem[van Dokkum et al. (2010)]{van Dokkum10} van Dokkum, P. G., et al. 2010, ApJ, 709, 1018

\bibitem[Zentner \& Bullock (2003)]{Zentner03} Zentner, A. \& Bullock, J.S.,  2003, ApJ, 598, 49

\end{thebibliography}
\end{document}